\documentclass[]{aa}
\usepackage[]{psfig}
\usepackage[]{txfonts}

\begin{document}

\title{The first CCD photometric study of the open cluster
NGC 2126}


\author{A. G\'asp\'ar\inst{1,5} \and L. L. Kiss\inst{2,5} \and 
T.R. Bedding\inst{2} \and A. Derekas\inst{2,5} \and S. Kaspi\inst{6} 
\and Cs. Kiss\inst{4}
\and K. S\'arneczky\inst{3,5} \and Gy.M. Szab\'o\inst{1} \and 
M. V\'aradi\inst{1,5}} 

\institute{Department of Experimental Physics and Astronomical Observatory,
University of Szeged,
Szeged, D\'om t\'er 9., H-6720 Hungary
\and School of Physics, University of Sydney 2006, Australia
\and Astronomical Observatory, Szeged, Hungary
\and Konkoly Observatory of the Hungarian Academy of Sciences, P.O. Box 67,
H-1525 Budapest, Hungary
\and Guest observer at Konkoly Observatory
\and School of Physics and Astronomy and the Wise Observatory,
Tel-Aviv University, Tel-Aviv 69978, Israel
}

\titlerunning{NGC 2126: physical parameters and variable stars}
\authorrunning{A. G\'asp\'ar et al.}
\offprints{L. L. Kiss,\\
e-mail: {\tt laszlo@physics.usyd.edu.au}}

\date{}

\abstract{We present the first CCD photometric observations of the northern
open cluster NGC~2126. Data were taken on eight nights in February and December
2002 with a total time span of $\sim$57 hours.  Almost 1000 individual $V$-band
frames were examined to find short-period variable stars. We discovered six new
variable stars, of which one is a promising candidate for an
eclipsing binary with a pulsating component. Two stars were classified
as $\delta$ Scuti stars and  one as Algol-type eclipsing binary. Two stars are
slow variables with ambiguous classification. From absolute $V(RI)_{\rm C}$ photometry we
have estimated the main characteristics of the cluster: 
$m-M=11\fm0\pm0\fm5,\ E(V-I)=0\fm4\pm0\fm1,\ E(V-R)=0\fm08\pm0\fm06\ 
(E(B-V)=0\fm2\pm0\fm15)$ and $d=1.3\pm0.6$ kpc. Cluster membership
is suggested for three variable stars from their positions on the
colour-magnitude diagram. 
\keywords{open cluster and associations: general -- open clusters 
and associations: individual: NGC~2126 -- stars: variables: general --
$\delta$ Sct}}

\maketitle

\section{Introduction}

Variable stars in clusters are crucial tools of stellar astrophysics:
the application of stellar evolutionary theories via isochrone fitting
of the colour-magnitude diagrams yields temperature, luminosity and 
age values for the member stars, which in turn draw strong constraints 
on pulsational properties (e.g. Frandsen \& Arentoft 1998) or 
binary evolution (Rucinski et al. 1996). Early research of $\delta$ Scuti
pulsation in open clusters (Breger 1972) and its revival with the wide
application of the CCD technique (for a review of the current state see  
Rodr\'\i guez \& Breger 2001) have recently been extended toward 
the new class of $\gamma$ Dor variables (Handler 1999), which have been 
found in a number of young and intermediate-age open clusters (Arentoft et al.
2001, Kim et al. 2001, Choo et al. 2003 and references therein). Clusters 
of different parameters can help map out the dependence of pulsational 
properties on age and metallicity, thus allowing better understanding of 
physical mechanisms driving the pulsation. On the other hand, detached 
eclipsing double-lined spectroscopic binaries can serve as very accurate 
distance indicators (e.g. Thompson et al. 2001), with the potential of 
feeding back to the isochrone fitting method itself. 
This work aims to contribute to these issues with the 
first CCD observations of the northern open cluster NGC~2126, with particular
emphasis on its variable star content.

NGC~2126 (= C0559+499, $\alpha_{2000}$=06$^h$02\fm55, 
$\delta_{2000}$=+49$^\circ$52$^\prime$, Tr\"umpler class III 2 m :b) is a
moderately rich typical galactic cluster with several dozens of members 
scattered in a region of 5--6$^\prime$ (Lyng\aa\ 1987) in the constellation
Auriga.  The only previous photometric study was presented by Cuffey
(1943), who made blue and red ($\lambda\lambda$~4300--6200)  photographic
observations. He estimated the distance to NGC~2126 as 950 pc, based on a
comparison to M~35. The WEBDA  catalog\footnote{\tt
http://obswww.unige.ch/webda} reflects this lack of data, as no other source is
listed for NGC~2126. 
This neglect turned our attention toward NGC~2126 and 
the present study presents an analysis of 8 nights of observations 
obtained in 2002. The measurements and data reduction are described in 
Sect.\ 2. Cluster parameters
and the six new variable stars are discussed in  Sect.\ 3, and a brief
summary is given in Sect.\ 4. 

\section{Observations and data reduction}

\begin{figure}
\begin{center}
\leavevmode
\psfig{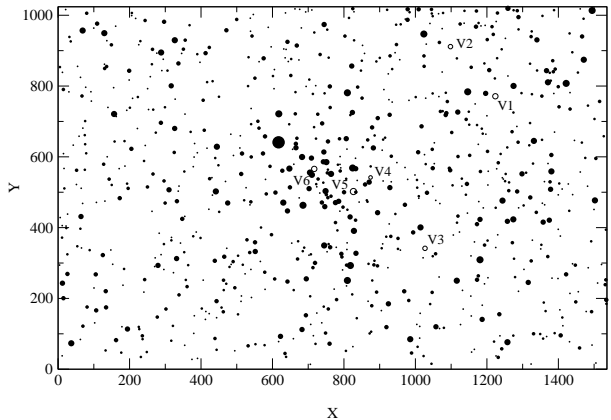}
\caption{The observed CCD field (29$^\prime\times18^\prime$) of the 
open cluster NGC~2126. The new variable stars are shown with open circles.
Symbol sizes are proportional to the stellar brightnesses. The largest
concentration of stars is about $8^\prime$ in diameter, centred at (750,550).
North is up, east is to the left.}
\label{terkep}
\end{center}
\end{figure}

CCD $V(RI)_{\rm C}$ observations were carried out on five consecutive nights in February 2002
and three nights in December 2002.
We used the 60/90/180 cm Schmidt telescope mounted at the Piszk\'estet\H{o}
Station of the Konkoly Observatory. The detector was a Photometrics AT-200 CCD
camera (1536$\times$1024 pixels, KAF-1600 chip with UV-coating). The image scale
was 1\farcs1/pixel, giving a 29$^\prime\times18^\prime$ field of view.
The observations consisted of $V$-band time-series observations 
(mostly 180 s exposures) and 
calibrated $V(RI)_{\rm C}$ photometric
observations (from 60 s to 600 s exposures).
The full observing log is presented in Table\ \ref{obs}. The observed field 
and the new variable stars are shown in Fig.\ \ref{terkep}.

\begin{table}
\begin{center}
\caption{The journal of observations.}
\label{obs}
\begin{tabular}{|lcccc|}
\hline
Date & $V$  & $R_{\rm C}$  & $I_{\rm C}$ & length\\
     & frames & frames &  frames & (hours)\\
\hline
2002 Feb. 1/2  & 83 & 2 & 3 & 8.5\\
2002 Feb. 2/3  & 108 & 1 & 1  & 8.8\\
2002 Feb. 3/4  & 142 & -- & -- & 8.7\\
2002 Feb. 4/5  & 134 & -- & -- & 8.7\\
2002 Feb. 5/6  & 129 & -- & -- & 8.7\\
2002 Dec. 23/24 & 62  & -- & -- & 3.4\\
2002 Dec. 25/26 & 215 & -- & -- & 5 \\
2002 Dec. 31/01 & 101 & -- & -- & 5 \\
\hline
Total: 8 nights & 974 & 3  & 4 & 56.8\\
\hline
\end{tabular}
\end{center}
\end{table}

The image reduction was done with standard tasks in IRAF\footnote
{IRAF is distributed by the National Optical Astronomy Observatories,
which are operated by the
Association of Universities for Research in Astronomy, Inc., under  
cooperative agreement with the National Science Foundation.}. For flat field
corrections, we used sky flat images taken during the evening and morning
twilights. We performed psf-photometry with the {\it daophot} package in IRAF
using the Moffatt point-spread function, which was found to be well-suited for 
fitting the slightly distorted stellar profiles of the used instrument 
(Kiss et al. 2001).

For photometric calibrations, we observed standard stars in the open
cluster M~67 from the list of Chevalier \& Ilovaisky (1991). The instrumental
magnitudes were transformed with the {\it photcal} package, according to 
the following standard transformation equations:

\begin{eqnarray}
V  &=&  v - 0.136 X + 0.126(V - R) - 5.180, \sigma=0\fm03 \\
 (V - I) &=& 1.028 ((v - i) - 0.079 X) + 0.348, \sigma=0\fm02 \\
(V - R) &=& 0.827 ((v - r) - 0.034 X) - 0.007, \sigma=0\fm04
\end{eqnarray}

\noindent where the symbols have their usual meanings. For the calibration 
we used 19 stars of the 29 listed by Chevalier \& Ilovaisky (1991). The colour
ranges were $-$0\fm03...0\fm64 and $-$0\fm07...1\fm21 for $V-R$ and $V-I$,
respectively, so that tranformations of the reddest stars in the field (about 
10\% of the sample) were extrapolations. Consequently, there might be some
systematic distorsion in the reddest part of the colour-magnitude diagram.

The transformation of the cluster was done in two steps. First, we
calibrated non-saturated stars on three images (one for each filter) confined
by consecutive M67 observations on February 2, 2002. Then we co-added a number of
$V$ band images and all $(RI)_{\rm C}$ frames and measured fainter stars
relative to previously calibrated local standards. We estimate the final 
photometric error as $\pm$0\fm05, which is likely to be
an overestimate for the brightest stars.

Time-series data were reduced in two separate subsets (5 nights in February 
and 3 nights in December), because the follow-up
observations in December, 2002 were obtained under a different instrumental
setup (the CCD imager was rotated by 90$^\circ$, so that a certain 
field of view was excluded from the measurements). Within the subsets,
CCD frames were matched by the task {\it imalign}. Since the cluster is 
not too crowded, instrumental magnitudes could be easily identified
based on 
the X-Y pixel coordinates. Stars lying closer than 30 pixels to the image
edge were rejected. Differential light curves were calculated with 
the ensemble normalization technique (Gilliland \& Brown 1988), for which 
dozens of stars of medium brightness were selected. We could identify almost
800 stars in the February run and that was the light curve dataset
in which we searched for variable stars.  The search was performed by
checking the light curve 
statistics and by visual inspection of the suspect time-series data.
We calculated the scatter of individual light curves for every star and 
plotted it against the apparent brightness. Outstanding points referred 
to either variable stars or constant stars with psf-fitting difficulties
(e.g. crowding). In this way we found six stars to be intrinsic variables.
In December, 2002, we focused on these six variables and the ensemble 
comparison stars, so that it is possible that there are unnoticed 
slow and/or long-period variables in the cluster field.

The light curves were further analysed to search for possible periodicities.
For this, we calculated Fourier spectra with Period98 (Sperl 1998). For the
eclipsing binaries, the phase dispersion minimisation (PDM, Stellingwerf 1978) 
provided a better period searching method. Our original data are available
electronically at the CDS via anonymous ftp to {\tt cdsarc.u-strasbg.fr
(130.79.128.5)} or via {\tt
http://cdsweb.u-strasbg.fr/cgi-bin/qcat?J/A+A/\\.../...}

\section{Results}

\subsection{The colour-magnitude diagram}

\begin{figure}
\begin{center}
\leavevmode
\psfig{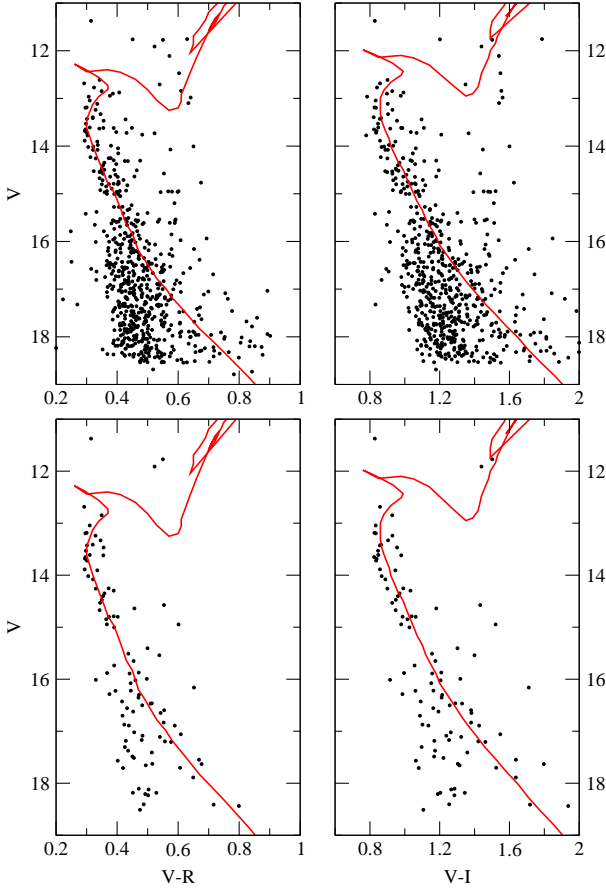}
\caption{The colour-magnitude diagrams of NGC~2126. Top row: full field of view;
bottom row: the central field of 8$^\prime$ diameter. The solid lines show 
the $\log t=9.1$ isochrone.}
\label{cmd}
\end{center}
\end{figure}

The physical parameters of the cluster were estimated by fitting isochrones
to the colour-magnitude diagram (CMD). In order to separate the cluster's stars from other field stars,
we have examined proper motions taken from the USNO B1.0  catalog (Monet et al.
2003). We found that of the 800 stars, almost 200 have proper motions detected 
in the USNO plate material. However, those stars are evenly distributed over the
field with no concentration around the cluster, so that NGC~2126 turned out to be
in the background. Consequently, we removed all stars from the CMD with non-zero
proper motions and the resulting CMDs are shown in the top row of 
Fig.\ \ref{cmd}. This procedure has been checked by two nearby galactic fields,
offset by about 1 degree north and south. A comparison of proper motion histograms
showed that we have indeed selected the galactic foreground.

In order to decrease background contamination, we kept only the inner 8$^\prime$
of the cluster for the isochrone fitting (bottom row in Fig.\ \ref{cmd}). This
region contains 103 stars between $V=11\fm3-18\fm5$, of which the majority seems
to be easily distinguishable from the background and some weak foreground. 
We assumed solar composition
and included reddening determination in the fitting procedure. The isochrones
were taken from Bertelli et al. (1994) and shifted individually to match the 
main sequence, turn-off point and red giant positions. We found that the overall
shape of the CMD is well reproduced with
isochrones of $\log t$ ranging from 9.0 to 9.3, $E(V-I)$ between 0\fm53 and
0\fm33, $E(V-R)$ between 0\fm14 and 0\fm03 and distance moduli between 11\fm5 and
10\fm4. We plot the $\log t=9.1$ isochrone in Fig.\ \ref{cmd}, showing the
``best'' fit, but the difference for the other isochrones is almost negligible.
Changing the composition makes the parameter ranges even wider, thus the
estimated parameter errors are quite large and a further spectroscopic study
addressing spectral types and chemical composition is desirable.

In this paper we adopt the following physical parameters: 
$m-M=11\fm0\pm0\fm6$, $E(V-I)=0\fm4\pm0\fm1$, 
$E(V-R)=0\fm08\pm0\fm06$. The reddenings were converted to 
$E(B-V)$ using the coefficients listed in Rieke \& Lebofsky (1985):
$E(V-R)/E(B-V)=0.78$, $E(V-I)/E(B-V)=1.60$. The result is 
$E(B-V)=0\fm2\pm0\fm15$, and the distance to the cluster 
(assuming $A_{\rm V}\approx3.1\times E(B-V)$) is $1.3\pm0.6$ kpc.

\subsection{New variable stars}

\begin{figure*}
\begin{center}
\leavevmode
\psfig{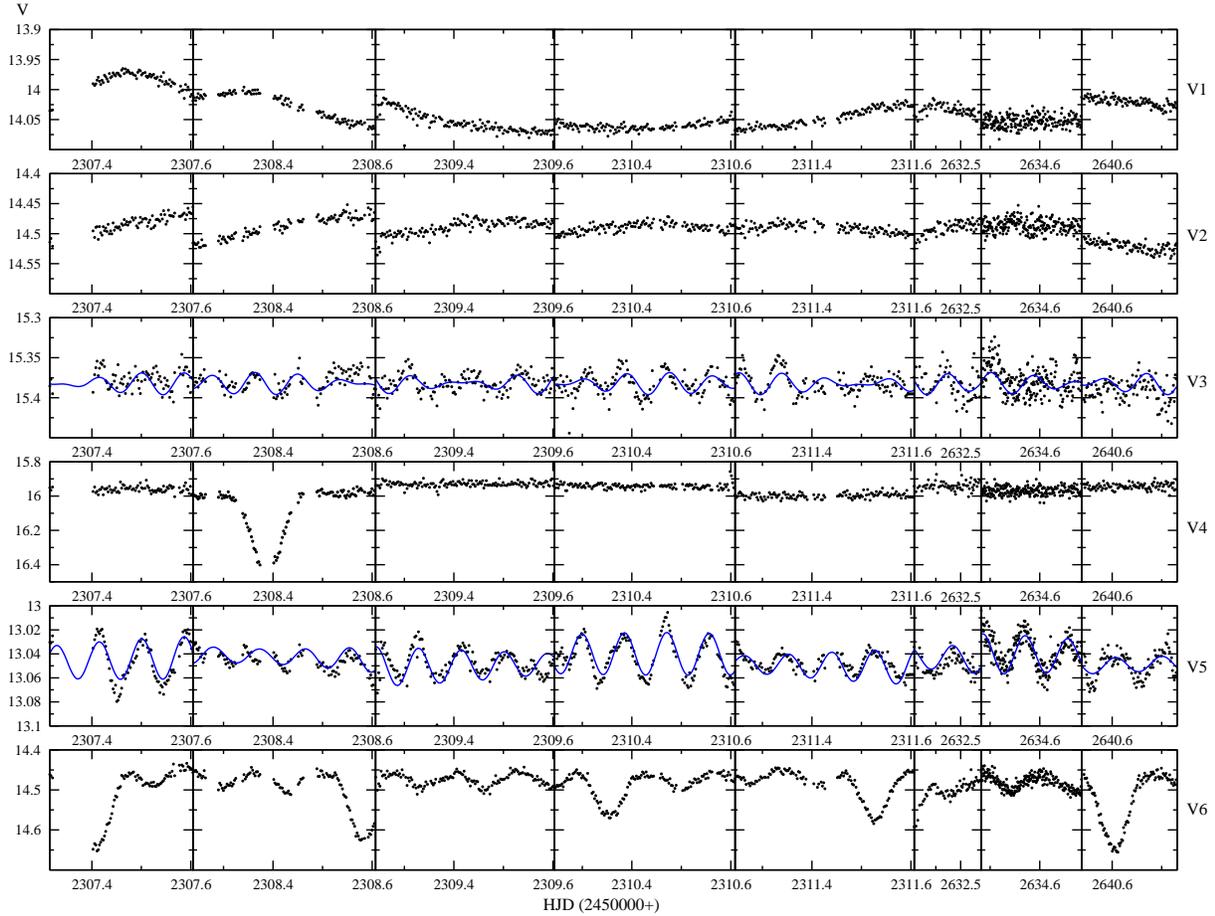}
\caption{Individual light curves of the six new variable stars. The solid
lines denote Fourier fits for V3 and V5.}
\label{fenygorbek}
\end{center}
\end{figure*}

Nightly light curves of the new variable stars are shown in Fig.\
\ref{fenygorbek} (numbered in order of increasing right ascension).
Differential light curve data were shifted so that the mean values
match the apparent magnitudes of the stars in the CMD\@.  Therefore, 
the absolute $V$ magnitudes in Fig.\ \ref{fenygorbek} are somewhat
uncertain, depending on the magnitude range of the
variables.

The light curve
shapes suggest the following classifications: V1 and V2 are slow variables of
ambiguous nature; V3 and V5 are  short-period variables, most
likely of $\delta$ Scuti type; V4 is an Algol-type  eclipsing binary with only
one observed minimum, while the light curve
of V6 is a mixture of $\delta$ Scuti-like oscillations and Algol-like eclipses.

\subsubsection{V1 and V2: slow variations}

\begin{figure}
\begin{center}
\leavevmode
\psfig{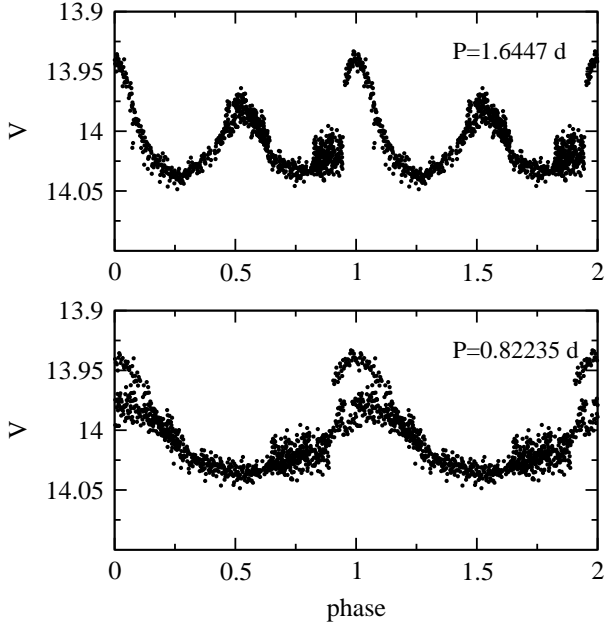}
\caption{Phase diagrams for V1 with two equally possible periods.}
\label{v1faz}
\end{center}
\end{figure}

Although these two stars showed clear variability, our dataset is
too short to determine reliable periods. For V1, both PDM and Fourier
analysis suggested P=1.64470 d or half that value, 0.82235 d (the 
uncertainties are a few in the last digit). We plot the
resulting phase diagrams in Fig.\ \ref{v1faz} (the epoch of maximum is
HJD 2452307.48).

With no colour light curve, the classification is uncertain. 
For 0.822235 d, one can speculate on possible $\gamma$ Dor-like oscillations.
These stars are early F-type main-sequence or subgiant stars with periods
0.4--3.0 d and $V$-band amplitudes less than 0\fm1, pulsating in 
non-radial gravity modes. According to the definition by Handler (1999), 
the $\gamma$ Dor instability strip covers the 7200--7700 K range on the
zero-age main sequence (ZAMS) and the 6900--7500 K range one magnitude
above the ZAMS. Assuming cluster membership, the dereddened colours of V1 are
$(V-R)_0=0\fm24$ and $(V-I)_0=0\fm48$. Standard tabulations  
give an effective temperature of 6500-7000 K (Cox 2000), which 
marginally supports
the classification as a $\gamma$~Dor star. Also, the cluster
membership would imply an absolute magnitude M$_{\rm V}$=3\fm0; 
the temperature and absolute magnitudes are consistent with 
the coolest $\gamma$ Dor variables.
In this case, the apparent amplitude modulation is due to the presence 
of other excited modes of pulsation and a proper asteroseismological analysis would
require a much longer dataset.

On the other hand, for P=1.6447 d, the double-peaked light curve 
is similar to those observed for spotted active binaries of RS~CVn 
subtype. For instance, recent data collection of Cutispoto et al. (2003) 
contains nice examples of very similar double-peaked curves of well-known 
active stars (BI~Cet, BC~Phe). Based on the present dataset, it is impossible
to distinguish between the two classifications. The fact that V1 lies
quite far from the cluster centre (about 10$^\prime$) weakens the possibility
of membership and  consequently makes the temperature and absolute
magnitude estimates less reliable.

For V2, the data can be equally well folded with periods around 0.5 d and 1
d, but neither period gives a continuous phase diagram with no
gap. Therefore, we cannot decide the cause of the low-amplitude variability
of this star (the total range is about 0\fm08).

\subsubsection{V4: an Algol-type binary}

\begin{figure}
\begin{center}
\leavevmode
\psfig{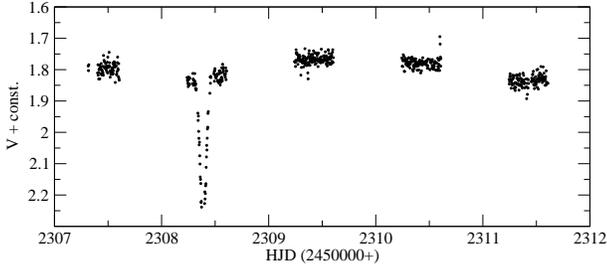}
\caption{The 2002 February data for V4.}
\label{v4lc}
\end{center}
\end{figure}

We observed only one minimum of the star (at HJD 2452308.387), 
so that no firm conclusion can
be drawn on its period. However, if we plot only the February subset in one 
light curve (Fig.\ \ref{v4lc}), the shape suggests a clearly 
visible reflection effect (about 0\fm1). 
Assuming symmetry around the unobserved secondary 
minimum, a tentative period of $\sim$3 days can be estimated (i.e. the next 
primary minimum happened immediately before or after the last night of 
observations in February).

\subsubsection{V3 and V5: multiply periodic oscillations}

\begin{figure}
\begin{center}
\leavevmode
\psfig{figure=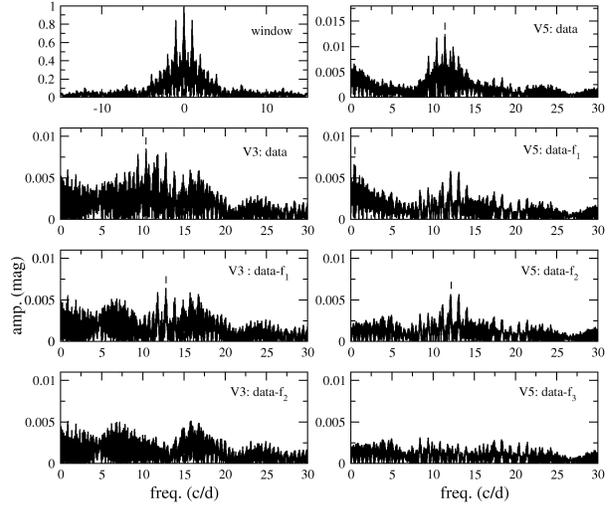,width=8cm}
\caption{Fourier analysis of V3 and V5.}
\label{v35fou}
\end{center}
\end{figure}

Both stars showed rapid oscillations with full amplitudes of a few tens
of millimags, characteristic of $\delta$ Scuti variables. We have determined
the pulsational frequencies with Fourier analysis consisting of iterative
prewhitening steps (see Fig.\ \ref{v35fou}). 

We could identify two frequencies
in both light curves (see Table\ \ref{v35frek}) with S/N ratios (Breger et al. 1993)
larger than 4. For V5, the low-frequency component refers to 
slight night-to-night variations of the mean brightness level, which we suspect
were caused by
a closely separated field star (see Fig.\ \ref{terkep}) lying  
$\sim7^{\prime\prime}$ from V5. It is fainter by about 4 mag, so that 
it may affect profile fitting to some small extent. Nightly 
variations of the seeing may also introduce variable contribution 
from this star.

In summary, V3 and V5 are probably multiply periodic pulsating 
$\delta$ Scuti variables. The frequency
ratios (V3: $f_1/f_2=0.81$, V5: $f_1/f_2=0.94$) suggest 
non-radial modes of pulsation for both stars. Small deviations from the 
light curve
fits (Fig.\ \ref{fenygorbek}) and residual structures of the 
prewhitened spectra (especially for V3) suggest the probable existence of more 
frequencies. Assuming their cluster membership implies 
the following dereddened colours and absolute magnitudes:
V3 --- $(V-R)_0=0\fm25$, $(V-I)_0=0\fm5$, $M_{\rm V}\approx4\fm4$;
V5 --- $(V-R)_0=0\fm23$, $(V-I)_0=0\fm43$, $M_{\rm V}\approx2\fm0$.
The latter values are consistent with typical parameters
of $\delta$ Sct stars (Rodr\'\i guez \& Breger 2001), 
we therefore conclude V5 is likely to be a member of the cluster. 
Moreover, its
proximity to the cluster centre supports this conclusion.
On the other hand, V3 seems to be a background object, since its absolute
magnitude, assuming cluster membership, places it well below the main
sequence.

\begin{table}
\begin{center}
\caption{The results of the period analyses.}
\label{v35frek}
\begin{tabular}{|llrrr|}
\hline
&i & f$_{\rm i}$ (d$^{-1}$) & amp. (mmag) & S/N\\
\hline
V3 &&  & & \\
&1  &  10.39   & 7.6 & 5\\
&2  &  12.79   & 6.5 & 4.3\\
\hline
V5& &  & & \\
&1  &  11.43   & 11.9 & 12\\
&2  &  0.42    & 5.3 & 3.8\\
&3  &  12.14   & 5.9 & 6\\
\hline
\end{tabular}
\end{center}
\end{table}

\subsubsection{V6: an eclipsing binary with a pulsating component?}

The most intriguing star is V6, which shows complex light variations. 
We observed well-defined minima -- three deeper and two shallower --
characteristic of an eclipsing
binary with components of different temperatures. Moreover, the star showed 
steady oscillations outside eclipses, with amplitude and cycle length
characteristic of $\delta$ Scuti-like pulsation (about 0\fm05 and 0.13 d).
The symmetric stellar profile, even on the best CCD images, excluded the 
possibility of two unresolved stars, so that we conclude V6 is an
eclipsing binary with at least one pulsating component. Such systems 
are spectacular targets for asteroseismology, because independent
determination of the physical parameters (mass, radius, temperature), applying
binary star astrophysics, gives strong constraints on the possible mode
identification (e.g. Mkrtichian et al. 2002, Kim et al. 2002a). 
In addition, eclipses of a pulsating star give spatial resolution across
the stellar surface and thereby provide the possibility for mode
identification (e.g. Reed et al. 2002). 

\begin{figure}
\begin{center}
\leavevmode
\psfig{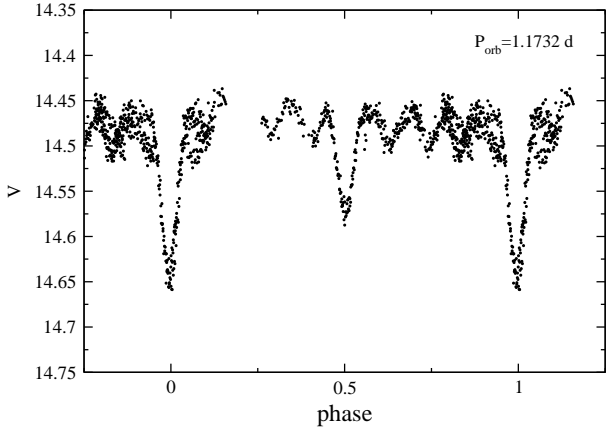}
\caption{Phase diagram for V6.}
\label{v6faz}
\end{center}
\end{figure}

The period analysis resulted in $P_{\rm orb}=1.17320(3)$ d. The data were folded
with this period and the phase diagram is shown in Fig.\ \ref{v6faz}. The 
depth is 0\fm2 and 0\fm12 for the primary and secondary minima, respectively.
The oscillations are clearly visible outside minima and we note
that the scatter of the phase diagram after folding only the February data 
was fairly small. This indicated that the oscillations were quite coherent 
within 5 consecutive nights in February. Their amplitude was about 0\fm05
associated with a period  $P_{\rm pul}\approx$0.13 d. 

\begin{figure}
\begin{center}
\leavevmode
\psfig{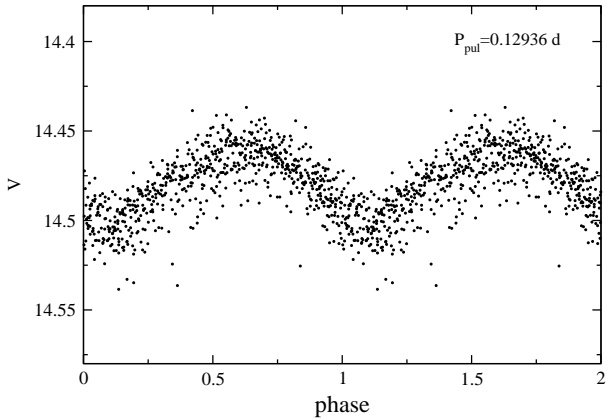}
\caption{Outside-eclipse data folded with $P_{\rm pul}=0.12936$ d.
The zero phase corresponds to a primary eclipse.}
\label{v6pulfaz}
\end{center}
\end{figure}

We also performed a separate period analysis of data from which eclipse
minima were excluded.  This showed the oscillations to be surprisingly
stable. They have a period $P_{\rm pul}=0.12936$ d and the phase diagram
shows remarkable small scatter (Fig.\ \ref{v6pulfaz}). The subsequent
prewhitening steps indicated a low-frequency component, approximately
2/T$_{\rm obs}$, and $f_3\approx f_{\rm pul}-f_{\rm orb}$. Another 
interesting result is that $f_{\rm pul}/f_{\rm
orb}=9.07$, suggesting that there might be a 1:9 resonance between the
orbital motion and pulsation.

\begin{figure}
\begin{center}
\leavevmode
\psfig{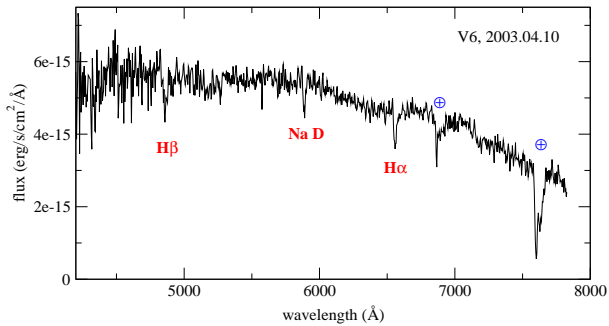}
\caption{A low-resolution spectrum obtained on April 10, 2003.}
\label{v6spec}
\end{center}
\end{figure}

We have also observed V6 with the 40$^{\prime\prime}$ telescope at 
the Tel-Aviv University Wise Observatory on 2003 April 10. The spectrograph 
10$^{\prime\prime}$ long-slit was aligned at PA=133.8$^\circ$ in order to 
maximize the separation between the
V6 spectrum and the nearby objects residing south-east to it. In this
way, we minimized the possibility of light from the nearby objects
contaminating the V6 spectrum. The spectrum shows prominent hydrogen absorption
lines accompanied by the sodium D-doublet (Fig.\ \ref{v6spec}). The spectral 
lines and the continuum shape suggest an F-type star, typical for a $\delta$ 
Scuti star, so that the oscillations may be attributed to $\delta$ Scuti-like
pulsations.

Out of the six variables, only V6 has 
non-zero proper motion ($\mu_\alpha \cos \delta=-14$ mas/year,
$\mu_\delta=10$ mas/year), thus cluster membership can be excluded.

In the literature, only a few similar system are reported, making V6 
a very interesting variable star. Rodr\'\i guez \& Breger (2001) listed 
nine $\delta$ Scuti variables in
eclipsing binaries, of which RZ~Cas (Ohshima et al. 2001) and AB~Cas
(Rodr\'\i guez et al. 1998) are the best studied. Compared to the known
examples, V6 is different in two important respects: {\it i)} the relative
amplitude ratio ($\Delta V_{\rm prim}/\Delta V_{\rm pul} \approx 4$) is
much smaller than that of any other similar stars (the next is V577~Oph,
with a ratio of 13); {\it ii)} the period ratio ($P_{\rm orb}/P_{\rm pul}
\approx 9$) is much smaller than usual -- from Table\ 3 in Rodr\'\i guez \&
Breger (2001), only WX~Eri has a ratio smaller than 20 (5.005).

Very recently, several new examples of pulsating components in eclipsing
binaries have been found (RX~Hya, Kim et al. 2002b, 2003; AB~Per, Kim et
al. 2002c, 2003; TW~Dra, Kim et al. 2003; AS~Eri, Gamarova et al. 2000), 
but all are typical low-amplitude
rapid pulsators not resembling V6.  In that sense, V6 is a unique candidate
for an eclipsing binary with a pulsating component. Future
spectrophotometric observations, accompanied by radial velocity
measurements, are needed to settle a number of questions regarding V6.  In
particular, are the cyclic changes caused by stellar pulsation? Which
component is pulsating? What is the true amplitude of the oscillations,
after removing the second light from the light curve (could this be the
first high-amplitude $\delta$ Scuti star in an eclipsing binary)?  Is there
indeed resonance between orbital motion and pulsation? If so, are the oscillations
induced by tidal forces (Willem \& Aerts 2002)?

\section{Conclusions}

\begin{figure}
\begin{center}
\leavevmode
\psfig{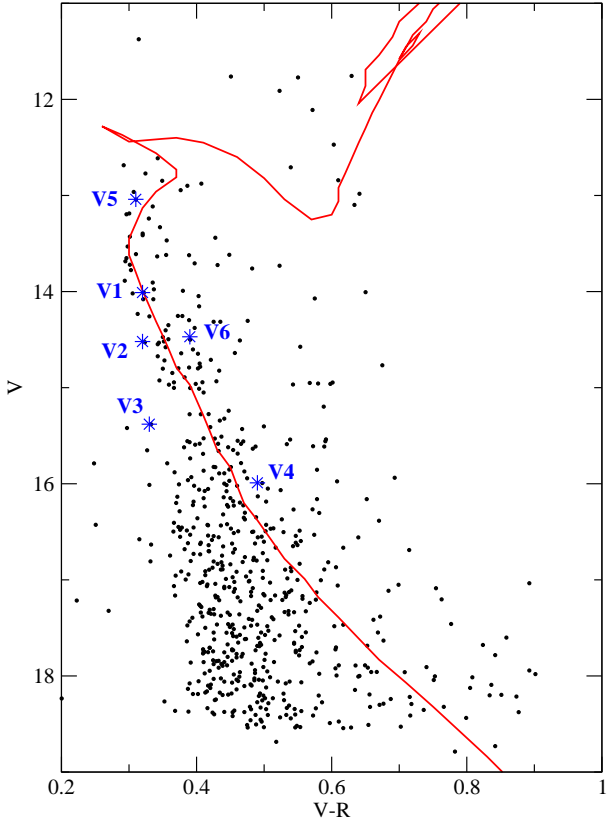}
\caption{Variable stars on the colour-magnitude diagram excluding foreground
stars (except V6).}
\label{varcmd}
\end{center}
\end{figure}

\begin{table*}
\begin{center}
\caption{The basic data of new variable stars. Except V6, coordinates and identifications 
were taken from the Guide Star Catalog, Version 2.2 (STSci 2001).}
\label{varadat}
{\scriptsize
\begin{tabular}{|c|lcccccccll|}
\hline
-- & GSC2.2 number& $\alpha_{2000}$  &$\delta_{2000}$ & $V$ & $(V-R)$ &$(V-I)$ & $P[^d]$ & $\Delta V$
& Epoch & Type\\
\hline\hline
V1 & N2110020273& $6^{h}01^{m}44.32^{s}$ & $+49^{o}56'32.5''$ & 14\fm01 & 0\fm32 & 0\fm88 & 1.6447 or 0.82235 &
0\fm12& 2452307.47 & GDOR? \\
V2 & N21100206826& $6^{h}01^{m}57.68^{s}$ & $+49^{o}58'56.5''$ & 14\fm52 & 0\fm32 & 0\fm88 & - & 0\fm08 & -- & --\\
V3 & N21100204078& $6^{h}02^{m}05,49^{s}$ & $+49^{o}49'14,5''$ & 15\fm38 & 0\fm33 &0\fm90 & 0.096, 0.078 & 0\fm08&
-- & DSCT\\
V4 & N21100204944& $6^{h}02^{m}21,25^{s}$ & $+49^{o}52'38,4''$ & 15\fm99 & 0\fm49 &1\fm31 & -- & 0\fm40& 2452308.387
& EA\\
V5 & N2110020321& $6^{h}02^{m}26,67^{s}$ & $+49^{o}51'55,2''$ & 13\fm04 & 0\fm31 &0\fm83 & 0.087, 0.082 & 0\fm05& --
& DSCT \\
V6 & --& $6^{h}02^{m}38,27^{s}$ & $+49^{o}53'04,7''$ & 14\fm47 & 0\fm39 &1\fm09 & 1.1732, 0.12936 & 0\fm23&
2452640.6072 & EA+DSCT:\\
\hline
\end{tabular}
}
\end{center}
\end{table*}

This paper presents the first CCD photometric study of the northern open cluster
NGC~2126. The $V(RI)_{\rm C}$ observations revealed the main characteristics
of the cluster and led to the discovery of six new variable stars.
The locations of the variables on the CMD (Fig.\ \ref{varcmd}) suggest 
membership for V1, V4 and V5; V2 and V3 are most likely non-member stars, 
while V6 is a foreground object. The most important results of this
research are the following:

\begin{enumerate}

\item We have  estimated important physical parameters of the cluster with 
standard photometric methods for the first time. 

\item We have discovered six variable stars in a field of $29^\prime \times 
18^\prime$ centred on the cluster, which is about $8^\prime$ in diameter.
Two are pulsating stars, one is an eclipsing binary with unknown period, one
showed a combined light variation of eclipses and pulsations, while two 
stars have ambiguous classification. 

\item The most interesting variable star is V6, which is a promising candidate 
for an eclipsing binary with a pulsating component. The period ratio 
$P_{\rm orb}/P_{\rm pul}=9.07$ may indicate a resonance between the orbital 
motion and pulsation.

\end{enumerate}

\noindent We summarize the basic data of the variable stars in Table\ \ref{varadat}.

\begin{acknowledgements}
This work has been supported by the FKFP Grant 0010/2001, 
OTKA Grants \#F043203, \#T032258 and the Australian 
Research Council. Astronomy at the Wise Observatory is supported by grants 
from the Israeli Academy of Sciences.
The NASA ADS Abstract Service was used to access data and references. This 
research has made use of the SIMBAD database, operated at CDS-Strasbourg, France.
\end{acknowledgements}

\end{document}